\documentclass[prl,twocolumn,superscriptaddress,showpacs,amsmath,amssymb]{revtex4}

\usepackage{graphicx}
\usepackage{dcolumn}
\usepackage{bm}
\usepackage{color}
\usepackage{mathrsfs}

\begin{document}

\preprint{APS/123-QED}

\title{Stable Skyrmions in $SU(2)$ Gauged Bose-Einstein Condensates}

\author{Takuto Kawakami}
\affiliation{Department of Physics, Okayama University, Okayama 700-8530 Japan}%
\author{Takeshi Mizushima}
\affiliation{Department of Physics, Okayama University, Okayama 700-8530 Japan}%
\author{Muneto Nitta}
\affiliation{Department of Physics, and Research and Education Center for Natural Sciences, Keio University, Kanagawa 223-8521 Japan}%
\author{Kazushige Machida}
\affiliation{Department of Physics, Okayama University, Okayama 700-8530 Japan}%
\date{\today}

\begin{abstract}
We demonstrate that the three-dimensional Skyrmion, which {has remained} elusive so far, spontaneously appears as the ground state of $SU$(2) symmetric Bose-Einstein condensates coupled with a non-Abelian gauge field. The gauge field is a three-dimensional analogue {of} {the} Rashba spin-orbit coupling. {Upon} squashing the $SO(3)$ symmetric gauge field to one- {or} two-dimensional shapes, we find that the ground state continuously undergoes {a} change from a three-dimensional to a one- {or} two-dimensional {Skyrmion}, which is identified by estimating winding numbers and helicity. All of {the} emerged Skyrmions {are physically} understandable with the concept of the helical modulation {in a unified way}. {These topological objects {might potentially be} realizable in two-component BECs experimentally.}
\end{abstract}

\pacs{03.75.Lm, 
      03.75.Mn, 
      67.85.Fg, 
      67.85.Jk}  

\maketitle

{\it Introduction.---}
The $n$-dimensional Skyrmions ($n\!\le\! 3$), classified by the $n$-th homotopy group ($\pi _n (S^n) \!=\! \mathbb{Z}$), have attracted much attention in various research fields{,} ranging from high-energy to condensed-matter physics~\cite{skyrme, brownbook, skyrmerev}.
It has been demonstrated that the two-dimensional ({2D}) Skyrmion spontaneously appears as the ground state in helical magnets MnSi and Fe$_{1-x}$Co$_x$Si~\cite{2dSK0,2dSK1,2dSK2}, a quantum Hall state~\cite{2dSK3}, {and continuous vortices analogous to Skyrmions in $^3$He-A~\cite{volovikbook,ichioka}}. Furthermore, it has recently been created by using the phase{-}imprinting technique in gaseous Bose-Einstein condensates (BECs)~\cite{choi}. {So far those are all {2D} objects.}

{The three-dimensional ({3D}) Skyrmion is a {particlelike} soliton in classical field theory, which was hypothetically introduced by Skyrme~{\cite{skyrme}} {to} describe baryons in a meson field. 
Although this hypothesis has attracted a lot of attentions for decades, 
{the} evidence has yet to be clarified.
One {difficulty} of the proposal is due to the {in}stability of Skyrmions:  
it is known that in the nonlinear $\sigma$-model, the gradient energy makes the {3D} Skyrmion unstable toward shrinkage, in spite of the topological stability~\cite{skyrme, skyrmerev} {because energetics differs from topology}. To prevent it from shrinkage, Skyrme~{\cite{skyrme}} added 
{\it by hand} the quartic differential term, which is the so-called {s}kyrme term. 
Instead of adding {an {\it ad hoc}} Skyrme term, 
it has been clarified that a non-Abelian gauge field in the form of
the t'Hooft-Polyakov monopole, which yields the scaling law {the} same as the {s}kyrme term, facilitates the stability of the {3D} Skyrmion~\cite{zachos}.}
It is, however, a {nonrealistic} and purely theoretical proposal, because such a non-Abelian gauge field does not exist in the meson field theory.

On the other hand, the stability of the {3D} Skyrmion{s} in {multicomponent} BECs ha{s} been investigated since their proposals~\cite{khawajanature, khawajapra}.
Although the schemes to create  and stabilize them~\cite{ruostekoski, savage, wuster, tokuno, oshikawa,battye} have theoretically been proposed, {the {3D} Skyrmion is still elusive both experimentally and theoretically. One reason is that the Skyrmions {in previous works} are merely {metastable} solutions of the energy functional~{\cite{ruostekoski, savage, wuster, tokuno, oshikawa,battye}}.}

The aim in this Letter is to clarify that {a} {3D} Skyrmion spontaneously emerges as the ``ground state'' of BECs,  coupled with a realistic non-Abelian gauge field, without the help of the Skyrme term. {This is the first proposal of the stable Skyrmions with the 3D {analogue} of Rashba SOCs.} First, we demonstrate that the stability of the {3D} Skyrmion is {physically} understandable with the concept of the helical modulation of the order parameter (OP)~\cite{wang,kawakami}. We show the phase diagram and the stable Skyrmion textures by numerically solving the full Gross-Pitaevskii (GP) equation. {Here, we {mainly} focus on a BEC with an $SU(2)$ symmetric interaction to capture the essential physics in the presence of the non-Abelian gauge field, 
as widely studied in the earlier works~\cite{ruostekoski, savage, wuster, tokuno, oshikawa,battye}. 
{T}he Hamiltonian is analogous to the Higgs sector of the Weinberg-Salam model of electroweak interactions~{\cite{gipson}}.}
{However, we also consider the stability of the 3D Skyrmion against the interaction without $SU(2)$ symmetry.}

Recently, there {was} a major breakthrough that {enabled} one to artificially imprint a gauge field in ultracold atoms~\cite{lin1,lin2,lin3}.
This is intriguing in the sense of accessibility to new topological phases and the appearance of spatially modulated ground states due to non-Abelian gauge fields~\cite{zhai}. The technique is based on the Raman coupling between hyperfine states of atoms which reconstructs the internal degrees of freedom to be degenerate pseudospin states. The Hamiltonian in the pseudospin representation has a fictitious Abelian or non-Abelian gauge {field} due to the adiabatic motion of {its} degenerate pseudospin states.

The schemes to generate {various} types of gauge fields have been proposed theoretically. In two-component BECs, there exist {methods} to generate {2D} and {3D} analogues to {the} Rashba spin-orbit coupling~\cite{juzeliunas, campbell, anderson} and a monopole field with a Dirac string~\cite{ruseckas}. {Recently,} the one-dimensional ({1D}) Rashba+Dresselhause type gauge field {was} experimentally realized by the NIST group~\cite{lin3}. {{Fermi gases coupled with a non-Abelian gauge field} {were} also investigated~\cite{vyasanakere, gong}.} In this Letter, we clarify how such non-Abelian gauge fields stabilize the {3D} Skyrmion.


{\it Phase diagram.---}
We start with the GP energy functional for two-component bosons with the {atom} mass $m$ and pseudospins $\uparrow,\downarrow$ as
\begin{eqnarray}\label{eq:gp}
	E=\mathcal{H}_0
	+ \int d^3{\bm r}\left[
	{\frac{1}{2}r^2n
	+c_0n^2+c_1n^2S_z^2} \right],
\end{eqnarray}
{where the particle density is $n(\bm{r})\!\equiv\!\Psi^\ast_\mu\Psi_\mu$ and the $\hat{\bm z}$-component of the local spin is $S_z(\bm{r})\!\equiv\!\frac{1}{2n}\Psi^\ast_\mu(\sigma_z)_{\mu\nu}\Psi_\nu$.}
{Throughout this {Letter}, the repeated Greek indices imply the sum on the spin $\mu, \nu, \eta \!=\! \uparrow, \downarrow$ {and we use the unit of $\hbar\!=\! m \!=\! \omega \!=\! 1$,} where $\omega$ is the trap frequency.}
{The $c_0\!+\!c_1$ is the interaction strength between bosons in the same pseudospin component and the $c_1$ indicates the interaction between the {intercomponents}.}
The OP $\Psi _{\mu}$ {for the condensate} obeys the normalization condition, $\int \Psi^{\ast}_{\mu}\Psi _{\mu} d^3{\bm r} \!=\! 1$. 
The single particle Hamiltonian $\mathcal{H}_0$ in Eq.~(\ref{eq:gp}) is given by
\begin{eqnarray}
	\mathcal{H}_0
	= \int d^3\bm{r} \left[ {\bm D}_{\mu\nu} \Psi _{\nu}(\bm{r}) \right]^{\dag}\cdot 
	\left[ {\bm D}_{\nu\eta} {\Psi}_{\eta}(\bm{r}) \right].
	\label{eq:H0}
\end{eqnarray}
Here, we set the covariant derivative ${\bm D}_{\mu\nu} \!=\! \left( -i{\bm \nabla}\sigma _0 + {\bm A} \right)_{\mu\nu}$ with the $2\times\!2$ unit matrix $\sigma _0$ and a non-Abelian gauge field $\bm{A}$.
{In most of this Letter, we set $c_1\!=\!0$ in Eq.~(\ref{eq:gp}), which ensures the $SU(2)$ symmetry of the interaction.
However, we will mention later that the finite $c_1$ does not make the Skyrmion unstable.}

\begin{figure}[tb]
		\includegraphics[width=85mm]{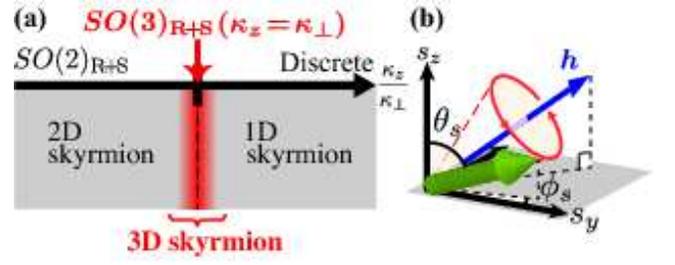}
		\caption{
				(Color online)
				(a) Schematic phase diagram of textures and the symmetries of $\mathcal{H}_0$.
				{At $\kappa_z\!=\!\kappa_\perp$, 
				the Hamiltonian has an $SO(3)_{R\!+\!S}$ symmetry.}
				(b) Schematic picture of the helical spin modulation along $\bm{h}$.
				{{The thick green (gray)} arrow depicts the local spin 
				$S_{j}^{(0)} \!\equiv\! \frac{1}{2}\{{\Psi}^{(0)}_{\mu}({\bm r}_0)\}^\ast(\sigma _{j})_{\mu\nu}{\Psi}_{\nu}^{(0)}({\bm r}_0)/n({\bm r}_0)$.}}
		\label{fig:helical}
\end{figure}

First of all, we summarize in Fig.~\ref{fig:helical}(a) the schematic phase diagram of {BECs} coupled with the {3D} non-Abelian {gauge} field,
\begin{eqnarray}\label{eq:3DRG}
	\bm{A} = \kappa_\perp \left(\sigma_x\hat{\bm{x}} + \sigma_y\hat{\bm{y}}\right) + \kappa_z\sigma_z\hat{\bm{z}},
\end{eqnarray}
where $\sigma _{j}$ denotes the $j$-th Pauli matrix. {This is {3D} analogue of a Rashba or Dresselhause type spin-orbit coupling which is known in the condensed matter context.} In fact, the scheme to generate this type of synthetic gauge field with {$\kappa_z\!=\!\kappa_\perp$~\cite{anderson} and $\kappa _z \!=\! 0$}~\cite{juzeliunas, campbell} {was} proposed theoretically, {and the case of $\kappa _{\perp}\!=\! 0$ was realized in the experiment~\cite{lin3}} {as mentioned}. In Fig.~\ref{fig:helical}(a), we find out the stable region of the {3D} Skyrmion texture, while the other regions are occupied by the {1D} {or} {2D} {Skyrmion}. It turns out that these textures {smoothly change} to each other. It is demonstrated in the rest {of this Letter} that all {types} of Skyrmions, whose textures are displayed in Fig.~\ref{fig:Skyrmion} are understandable with the helical modulation of the OP which can be parameterized on the three-dimensional surface $S^3$ {in the OP space}.

{\it Helical spin modulation and Skyrmions.---}
The concept of the helical modulation of the OP \cite{kawakami} provides a good {starting point} to the understanding of the stability and {smooth transition} of Skyrmions under the non-Abelian gauge field described in Eq.~(\ref{eq:3DRG}).

The OP of {a two-component BEC} is {always} parameterized, the $U(1)$ phase {$\Phi\!\equiv\!\Phi(\bm{r})$, {and} {the angles} $\phi_\mathrm{s}\!\equiv\!\phi _\mathrm{s}(\bm{r})$ {and} $\theta_\mathrm{s}\!\equiv\!\theta _\mathrm{s}(\bm{r})$,} as
\begin{eqnarray}\label{eq:op}
	\left(\begin{array}{c}
	\Psi_\uparrow(\bm{r})\\
	\Psi_\downarrow(\bm{r})
	\end{array}\right)
	=
	\sqrt{n(\bm{r})}e^{i\Phi}e^{-i\sigma_z\phi_\mathrm{s}/2}e^{-i\sigma_y\theta_\mathrm{s}/2}
	\left(\begin{array}{c}
	1\\
	0
	\end{array}\right).
\end{eqnarray}
Here, as shown in Fig.~\ref{fig:helical}(b), $\phi _\mathrm{s}(\bm{r})$ and $\theta _\mathrm{s}(\bm{r})$ denote the direction of the local spin, $S_{j}({\bm r}) \!\equiv\! \frac{1}{2}{\Psi}^{\ast}_{\mu}({\bm r})(\sigma _{j})_{\mu\nu}{ \Psi}_{\nu}({\bm r})/n({\bm r})$.
Since the OP manifold in Eq.~(\ref{eq:op}) is mapped onto a three-dimensional surface $SU(2)\!\simeq\!S^3$, 
two-component BECs have a topological object classified by the homotopy group $\pi _3(S^3)\!=\!\mathbb{Z}$.

Turning now to a non-Abelian gauged BEC, we demonstrate that the helical modulation of the OP due to the field ${\bm A}$ makes the Skyrmion stable even without the Skyrme term. {It is crucial to observe that} the helical modulation with an arbitrary {modulation vector} ${\bm h}$ {can be written down} within the $SU(2)$ symmetric OP in Eq.~(\ref{eq:op}) as
\begin{eqnarray}
	{\Psi}^{\mathrm{H}}_\mu(\bm{r},\bm{h})\!=\!\mathcal{U}_{\mu\nu}(\hat{\bm h}, 2\bm{h}\cdot\bm{r}){\Psi}^{(0)}_\nu(\bm{r}_0), 
	\label{eq:helical}
\end{eqnarray}
where the $2\!\times\!2$ matrix $\mathcal{U}(\hat{\bm n},\varphi)\!{=\!\exp[i(\varphi/2)(\hat{\bm n}\!\cdot\!\bm{\sigma})] }\!\in\! SU(2)$ denotes the $SU(2)$ rotation of an arbitrary OP ${\Psi}^{(0)}_\mu$ around $\hat{\bm n}$ by the angle $\varphi$. 
Figure~\ref{fig:helical}{(b)} shows the schematic picture of this modulation.
The helicity originates from the single particle spectrum $E_0\!=\! \mathcal{H}_0$ of ideal Bose gases in the thermodynamic limit {$\Psi_\mu(\bm{r})\!=\!e^{i\bm{k}\cdot\bm{r}}\Psi_\mu^{(0)}$}, which is given by
	$E_0 = k^2+\kappa^2\pm 2\sqrt{\kappa_z^2 k_z^2 + \kappa_\perp^2 (k_x^2+k_y^2)}$
{with $\kappa^2\!=\!\kappa_z^2\!+\!2\kappa_\perp^2$}. It turns out that the OP of the ground state is spatially modulated, since $E_0$ has {minima} on the finite ${\bm k}$. {Then,} the helical modulation with {$\hat{\bm h}$ in Eq.~(\ref{eq:helical})} is {the} superposition of momentum {eigenstates} $\bm{k}\!\parallel\!\pm\bm{h}$. {T}he spatial inversion symmetry of $\mathcal{H}_0$ guarantees the degeneracy of the helical modulation starting from {an arbitrary} $\Psi^{(0)}_\mu$.



For $\kappa_z\!=\!\kappa_\perp\!=\!\kappa{/\sqrt{3}}$, the single particle Hamiltonian $\mathcal{H}_0$ in Eq.~(\ref{eq:gp}) is invariant under the simultaneous rotation of spin and real spaces $SO(3)_{{\rm R\!+\!S}}$. {Since} $E_0$ has {minima on} surface $k \!=\! \kappa{/\sqrt{3}}${,} the helical spin modulation in Eq.~(\ref{eq:helical}) is degenerate for any direction of $\hat{\bm h}\!\parallel\! \hat{\bm k}$. Note that the {{3D} helical modulation} ${\Psi}^{\mathrm{H}}_\mu({\bm r},{\bm{h}\parallel{\bm r}})$ propagating with all the directions of $\bm{h}$ along $\bm{r}$ fulfills {the OP manifold $S^3$} within $0\!\le\!{{\bm h}\cdot{\bm r}}\!\le\!\pi$. Thus, this texture {${\Psi}^{\mathrm{3D}}_\mu\!\equiv\!{\Psi}^{\mathrm{H}}_\mu(\hat{\bm r},2{{\bm h}\cdot{\bm r}})$} is the {3D} Skyrmion, which is the candidate of the ground state.

In contrast, for the {region} $\kappa_z/\kappa_\perp\!<\!1$ in Fig.~\ref{fig:helical}(a), the $SO(3)_{\rm R\!+\!S}$ symmetry in $\mathcal{H}_0$ {in} Eq.~(\ref{eq:gp}) is broken into the $SO(2)_{\rm R \!+\!S}$ that denotes the joint rotation of spin and real spaces around the $\hat{\bm z}$-axis. $E_0$ also has a minimum line along $k_{\perp} \!\equiv\! \sqrt{k^2_x+k^2_y} \!=\! \kappa _{\perp}$. Therefore, it turns out that the possible stable texture for $\kappa_z\!<\!\kappa_\perp$ is the radial or the {1D} helical spin modulation {expressed by Eq.~(\ref{eq:helical}) with ${\bm h} \!\parallel\!(k_x,k_y,0)$}, where the former corresponds to {a} {2D} Skyrmion~\cite{kasamatsu}.

For {the region} $\kappa_z/\kappa_\perp\!>\!1$ in Fig.~\ref{fig:helical}(a), the Hamiltonian (\ref{eq:gp}) still remains invariant under the discrete symmetry that is the simultaneous $\pi$-rotation of spin and real spaces around the $\hat{\bm x}$- or $\hat{\bm y}$-axis, where $E_0$ has minima at ${\bm k} \!=\! \pm \kappa _z \hat{\bm z}$.
Then the most stable modulation vector $\bm{h}$ is confined to the $\hat{\bm z}$-axis, implying that the possible stable texture is the {1D} helical modulation along $\hat{\bm z}$-axis.
Since this spin-rotation along the $\hat{\bm z}$-axis consists of the $U(1)$ degrees of freedom, we can interpret this helical spin modulation as a {1D} Skyrmion.

{\it Stable Skyrmion textures.---}
In order to quantitatively discuss the candidates of the {stable spin} textures, 
we numerically minimize the full Gross-Pitaevskii energy functional (\ref{eq:gp}) {by using the imaginary time evolution scheme~\cite{mizushima} with a spatial grid of 121$^3$}.
{We have verified that the obtained solution is the true ground state by starting the calculation from a variety of initial condition{s} {such as the uniform spin, helical modulation with a variety of modulation vectors, {and} 2D {and} 3D Skyrmions}.}
For numerical calculations, we use {$c_0 \!=\!100$}, where the density profile yields a {Gaussian}-like shape.
{This parameter corresponds to the trap frequency $\omega/2\pi\!=\!100$ Hz and $s$-wave scattering length of the same component $a_{11}\!=\!a_{22}\!=\!5$ nm with the $^{87}$Rb mass and number of trapped atoms $N\!=\!5\!\times \!10^4$. These values define the length of the trap unit as $\sqrt{\hbar/m \omega}\!\simeq\! 1 $ $\mu$m.}
{To quantify the size of the atomic cloud, we use the Thomas-Fermi radius {$R_0\!=\!\sqrt{2\mu_0}\!=\!2.99$ in the trap unit, where $\mu_0\!=\!(15c_0/2\pi)^{2/5}/2$ is the chemical potential with the Thomas-Fermi approximation.} 

\begin{figure}[tb]
		\includegraphics[width=85mm]{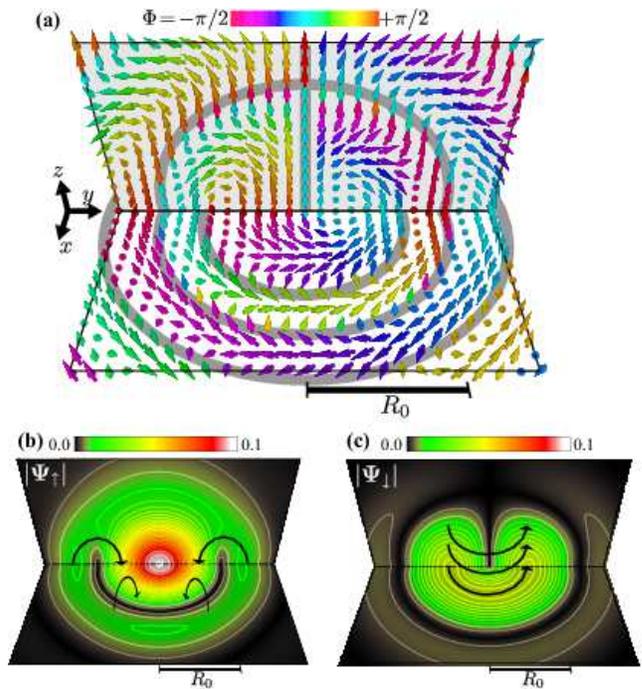}
		\caption{
				(Color online)
				The spatial profile of the stable 3D Skyrmion in the $\hat{\bm x}$-$\hat{\bm y}$
				and $\hat{\bm z}$-$\hat{\bm x}$ planes.
				The arrows and {their} color{s} in (a) indicate the pseudospin direction and the $U(1)$ phase of the OP{, respectively}.
				The gray line{s} in (a) imply the singularity of $\Phi$ and $\phi_\mathrm{s}$. 
				The color maps of (b) and (c) give the amplitudes $|\Psi_\uparrow(\bm{r})|$ and 
				$|\Psi_\downarrow(\bm{r})|$, respectively.
				The black arrows in (b) and (c) denote the directions of the phase winding of each component.
				These results are obtained {with} the parameter 
				$(\kappa_\perp R_0,\ \kappa_z R_0)\!=\!(5.07,5.07),$ {where $R_0\!=\!(15c_0/2\pi)^{1/5}$}.
				}
		\label{fig:Skyrmion}
\end{figure}

{Figure~\ref{fig:Skyrmion} shows the stable {3D Skyrmion} obtained from the numerical minimization of Eq.~(\ref{eq:gp}) {in the presence of the gauge field with $\kappa_\perp\!=\!\kappa_z$ and the spherical trap potential.} Figure.~\ref{fig:Skyrmion}(a) shows that the pseudospin texture helically modulates along {the} 3D radial direction. In this texture, the singularity of {the $U(1)$ phase} $\Phi$ and {the angle} $\phi_\mathrm{s}$ {introduced in Eq.~(\ref{eq:op})} exists on the gray line where the local spin points to {$\hat{\bm z}$} ($\Psi_\downarrow$-singularity) or {$-\hat{\bm z}$} ($\Psi_\uparrow$-singularity) direction.
As shown in Fig.~\ref{fig:Skyrmion}(b) and \ref{fig:Skyrmion}(c), OP component $\Psi_\mu$ accumulates the phase $2\pi$ on the path enclosing its singularity.
Namely, $\Psi_\uparrow$ {forms} the vortex ring and its ring-singularity is fulfilled by $\Psi_\downarrow$ with the phase winding.
This can be interpreted as the so-called ``vorton'' structure known {in {high-energy} physics}~\cite{davis,volovikbook}.}

{The 3D Skyrmion is identified by the winding number $\pi_3(S^3)$ of the map from real space to the OP manifold $SU(2)\!\simeq\!S^3$. The winding number defined by Refs.~\cite{skyrme, battye, ruostekoski, savage, wuster}}
\begin{eqnarray}
W_\mathrm{3D}=\frac{1}{8\pi^2}\int d^3\bm{r} \epsilon_{ijk} \sin\theta_{S} (\partial_i \theta_{S})(\partial_j\phi_{S})(\partial_k\Phi)
\label{eq:W3D}
\end{eqnarray}
{counts how many times the map warps the OP manifold.}
Figure~\ref{fig:winding}(a) shows the winding number $W_{\rm 3D}$ in the plane of $\kappa _z/\kappa _{\perp}$ and {$R_{\rm 0}\kappa _{\perp}$}, obtained from the numerical solution of the full GP equation. Here, we estimate Eq.~(\ref{eq:W3D}) in the region of $r\!\le\!1.5R_0$. It is seen from Fig.~\ref{fig:winding}(a) that $W_\mathrm{3D}$ increases with growth of the $\kappa_\perp/ R_0$ near the $\kappa_z/\kappa_\perp\!=\!1$ line. Hence, in this region the {3D} Skyrmion becomes stable. 
This continuous increase of $W_{\rm 3D}$ is because of the absence of the boundary condition that ${\Psi}_\mu(\bm{r}\!\rightarrow\!\infty)$ is {nonzero} and uniform, {assumed for Skyrmions in other contexts}. So far as the {boundary} condition is satisfied, $W_\mathrm{3D}\!\in\!\mathbb{Z}$ and {3D} Skyrmion is topologically stable. 
However, the presence of the vector potential $\bm{A}$ makes ${\Psi}_\mu(\bm{r}\!\rightarrow\!\infty)$ {nonuniform} even {if} we ignore the effect of the trap. 
In fact, with increasing of $\kappa_z\!=\!\kappa_\perp {\!\gtrsim\!R_0^{-1}}$, the characteristic length of the helical spin modulation becomes smaller and the shell-shaped Skyrmion penetrates from outside of the system in addition to a unit of the Skyrmion. Therefore, as $\kappa$ increases, $W_{\rm 3D}$ increases continuously and monotonically {as shown} in Fig.~\ref{fig:winding}(a).

\begin{figure}[tb]
		\includegraphics[width=85mm]{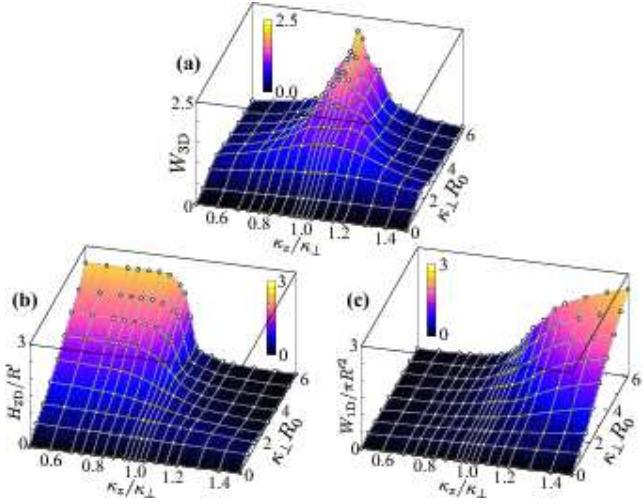}
		\caption{
				(Color online)
				Stereographic plots of $W_\mathrm{3D}$ (a), $H_\mathrm{2D}/R'$ (b), and $W_\mathrm{1D}/\pi R^{\prime 2}$ (c) 
				in the plane of $\kappa_z/\kappa_\perp$ and $\kappa_\perp R_0$.
				These denote the winding numbers and helicity which quantify the 1D, 2D, and {3D} Skyrmions. 
				We estimate these values in the region of $r\!\le\!1.5R_0\!\equiv\!R'$.
				}
		\label{fig:winding}
\end{figure}

{\it One- and two-dimensional Skyrmions.---}
{The 2D Skyrmion texture is described as {${\Psi}^{\mathrm{2D}}_{\mu}\!=\!{\Psi}^{\mathrm{H}}_\mu({\bm r},{\bm{h}\parallel{\bm \rho}})$, where $\bm{\rho}\!=\!{\bm x}\!+\! {\bm y}$}.}
{The Skyrmion ${\Psi}^{\mathrm{2D}}_{\mu}$ consists of the helical spin modulation along the 2D radial direction, which is favored by the gauge field described in Eq.~(\ref{eq:3DRG}) as mentioned above.}
{This texture occupies the {region} $\kappa_z/\kappa_\perp\!\lesssim\!1$ in Fig.~\ref{fig:helical}(a).} 
If we ignore the $U(1)$ phase $\Phi$, the winding number of this {2D} Skyrmion can be defined as 
$W_\mathrm{2D}\!=\!\frac{1}{4\pi}\int d^3\bm{r} \epsilon_{ij} \sin\theta_{S} (\partial_i\theta_{S})(\partial_j\phi_{S})$~\cite{kasamatsu}.

We should notice that the winding number of ${\Psi}^{\mathrm{2D}}_{\mu}$ is determined by the domain of the rotating angle $\Omega$;
$W_\mathrm{2D}\!=\!+1$ ($-1$) is accumulated within $n\pi\!<\!\Omega\!<\!(n\!+\!1/2)\pi$ ($(n\!-\!1/2)\pi\!<\!\Omega\!<\!n\pi$).
{This implies that} with increasing $\kappa_\perp$, the Skyrmion ($W_\mathrm{2D}\!=\! + 1$) and anti-Skyrmion ($W_\mathrm{2D}\!=\! - 1$) penetrates from outside alternately. Hence, $W_{\rm 2D}$ {of 2D Skyrmion ${\Psi}^{\mathrm{2D}}_{\mu}$ oscillates with increasing $\kappa _{\perp}$.} Instead of $W_{\rm 2D}$, one can estimate the size of the {2D} Skyrmion with the helicity {introduced as}
	{$H_\mathrm{2D}\!=\!\frac{1}{4\pi}\int d^3\bm{r} \left|\epsilon_{ij} \sin\theta_{S} (\partial_i\theta_{S})(\partial_j\phi_{S})\right|$}.

{The {1D} Skyrmion along the $\hat{\bm z}$-axis is described as ${\Psi}^{\mathrm{1D}}_{\mu}\!=\!{\Psi}^{\mathrm{H}}_\mu({\bm r},{\bm{h}\parallel{\bm z}})$. This appears as the ground state in the {region} $\kappa_z/\kappa_\perp\!\gtrsim\!1$ of Fig.~\ref{fig:helical}(a), where the modulation vector ${\bm h}\!\propto\! \pm \kappa_z \hat{\bm z}$ is favored as discussed above.} The winding number for the {1D} Skyrmion is introduced as 
	{$W_\mathrm{1D}\!=\!\frac{1}{2\pi}\int d{z} 
	\frac{{\Psi}^{\ast}_{\mu}\partial_z \left( \sigma_z \right) _{\mu\nu} \Psi _{\nu}}{{\Psi}^{\ast}_{\eta}{\Psi}_{\eta}}
	+\mathrm{c.c.}$}.
This corresponds to the phase accumulation which quantifies the spin current of $S_z$ along $\hat{\bm z}$-axis.

We plot $W_{\rm 1D}$ and $H_{\rm 2D}$ in Figs.~\ref{fig:winding}(b) and (c), where they reveal the stable region of the {1D} and {2D} Skyrmions in the parameter space spanned by $\kappa _z$ and $\kappa _{\perp}$. In the {region} $\kappa_z/\kappa_\perp\!\lesssim\!1$, the {2D} Skyrmion is stabilized where the helicity $H_\mathrm{2D}$ increases and $W_\mathrm{3D}$ and $W_\mathrm{1D}$ are suppressed. In the other {region}, $\kappa_z/\kappa_\perp\!\gtrsim\!1$, the growth of $W_\mathrm{1D}$ and the suppression of $W_\mathrm{3D}$ and $H_\mathrm{2D}$ indicate the appearance of the {1D} Skyrmion. Figure~\ref{fig:winding} provides evidence of the schematic phase diagram in Fig.~\ref{fig:helical}(a). 

Three types of Skyrmions can be continuously transformed {into} each other and the transition between them is identified as the second-order transition.
The cylindrical singularity of the component $\Psi_\uparrow$ and $\Psi_\downarrow$ in the {2D} Skyrmion 
continuously deforms to sphere and ring{-}shaped ones and then the texture changes into the {3D} Skyrmion. This is confirmed by the imaginary time evolution of the GP functional (\ref{eq:gp}), which demonstrates that the {2D} Skyrmion state becomes unstable in the vicinity of $\kappa _z/\kappa _{\perp} \!\sim \! 1$ toward the {3D} Skyrmion.
The {3D} Skyrmion state also becomes unstable in the region of $\kappa _z /\kappa _{\perp} \!\gtrsim\! 1$, where the singular line and sphere of $\Psi_\uparrow$ described in Fig.~\ref{fig:Skyrmion}(a). 
Hence, the {2D} Skyrmion solution smoothly transforms to {1D} Skyrmion through the {3D} Skyrmion as $\kappa_z/\kappa_\perp$ varies from $+0$ to $+\infty$, 
{and} the {3D} Skyrmion texture spontaneously appears at $\kappa_z/\kappa_\perp\!\simeq\!1$.

{{We emphasize that 
the {smooth transition} behavior between Skyrmions ensures the stability region of the {3D} Skyrmion against {finite $c_1\!\neq\!0${,} which breaks the $SU(2)$ symmetry in Eq.~\ref{eq:gp}.}  This is because the stability of {1D} and {2D} Skyrmions in the limit $\kappa_z\!\rightarrow\!0$ and $\kappa_\perp\!\rightarrow\!0$ stays unchanged in $c_1\!\neq\!0$~\cite{wang,hu}. In fact, within $0\!<\!c_1/c_0\!\lesssim\!1$, we numerically confirm that the winding number $W_\mathrm{3D}$ increases as approaching $\kappa_z/\kappa_\perp\!\rightarrow\!1$  {and the 3D Skyrmion ${\Psi}^{\mathrm{3D}}\!=\!\mathcal{U}_{\mu\nu}(\hat{\bm r}, 2\kappa r){\Psi}_\mu^{(0)}$ with $({\Psi}_\uparrow^{(0)},{\Psi}_\downarrow^{(0)})\!=\!(1,0)$ is stabilized. In the case of $c_1/c_0\!<\!0$, the 3D Skyrmion with $({\Psi}_\uparrow^{(0)},{\Psi}_\downarrow^{(0)})\!=\!(1,1)$ is also stabilized.}}}

{\it Skyrmions in other systems.---}
The stability of the {3D} Skyrmion in other spinor systems is worth a mention in passing. For instance, we find that the {3D} gauge field in Eq.~(\ref{eq:3DRG}) {cannot} stabilize the {3D} Skyrmion in the polar phase of a hyperfine spin $F\!=\!1$ spinor BEC, which is the ``knot'' soliton. This is because the helical modulation of the polar OP is not degenerate for the {3D} direction of the modulation vector~\cite{babaev,kawaguchi}. In fact, the OP manifold in the polar phase reduces to {an} $SO(2)$ symmetry in spin space. Hence, since the helical modulation vector along the direction of the $SO(2)$ rotation axis {cannot} gain the single particle energy $E_0$, only the {2D} or {1D} Skyrmion can be stable in this system. {In the ferromagnetic phase of an $F\!=\!1$ spinor BEC, the 1D Skyrmion~\cite{wang} and the 2D half-Skyrmion~\cite{su} are proposed. However, the problem on the stability of the 3D Skyrmion remains as nontrivial.}


{\it Conclusions.---}
{
Here, we have demonstrated that the 3D Skyrmion spontaneously appears as the ground state of two-component BECs coupled with a non-Abelian gauge field{, which is a} 3D analogue {of} {the} Rashba spin-orbit coupling. {The appropriate gauge field and the spherical density distribution due to the trap potential are necessary to stabilize the Skyrmions.} {Upon} squashing the 3D gauge field to 1D {or} 2D {shape}, the 3D Skyrmion continuously undergoes {a} change to 1D {or} 2D {Skyrmion}. We computed the ground state phase diagram with the winding numbers and helicity, which is covered by three types of Skyrmions. All of {the} emerged Skyrmions {are} understandable with the concept of the helical modulation of the order parameter. {In addition to the $SU(2)$ symmetric interaction, we also consider the stability of Skyrmions against an $SU(2)$ nonsymmetric interaction. Then, we confirmed that the asymmetric term does not alter the phase diagram. 
However, the interaction in the presence of a synthetic gauge field depends on the detail of the scheme to imprint it. The complete phase diagram in the more realistic situation remains as a future problem.}
}

The authors thank {Sankalpa Ghosh} and M. Ichioka for helpful discussions.
This work was supported by JSPS (No. 2200247703, 2074023303, 23740198, 2134010303) and the Topological Quantum Phenomena (No. 22103005, 23103515) KAKENHI on Innovation areas from MEXT.

{\it Note added.---}
After we submitted this paper, we became aware of two papers~\cite{li,anderson2} which discuss the ground state properties under the same gauge field.

\end{document}